\documentclass{emulateapj}

\usepackage{amsmath} 
\usepackage{amssymb}  
\usepackage{amsthm}
\usepackage{color}
\usepackage{url}
\usepackage{bm}  

\newcommand{\tamas}[1]{\textcolor{red}{#1 [Tamas]}}

\newcommand{\be}{\begin{equation}}
\newcommand{\ee}{\end{equation}}

\newcommand{\cfig}[1]{Figure~\ref{#1}}
\newcommand{\csect}[1]{Section~\ref{#1}}

\renewcommand{\O}[1]{\mbox{$\mathcal{O}(#1)$}}

\newcommand{\obs}{\bm{x}} 
\newcommand{\like}{\mathcal{L}}  
\newcommand{\mlike}{\mathcal{M}}
\newcommand{\drxn}{\bm{\omega}}  
\newcommand{\slike}{\ell}  
\newcommand{\sset}{S}  
\newcommand{\cset}{\mathcal{C}}  
\newcommand{\partn}{\mathcal{P}}  

\newtheorem{theorem}{Theorem}[section]

\DeclareMathOperator{\Tr}{Tr}

\begin{document}

\title{Probabilistic Cross-Identification in Crowded Fields as an Assignment Problem}

\author{
Tam\'{a}s Budav\'{a}ri\altaffilmark{1,2,3}
and
Amitabh Basu\altaffilmark{1}}

\altaffiltext{1}{Dept.\ of Applied Mathematics \& Statistics, Johns Hopkins University, 3400 N. Charles St., MD 21218, USA}
\altaffiltext{2}{Dept.\ of Computer Science, Johns Hopkins University,  3400 N. Charles St., MD 21218, USA}
\altaffiltext{3}{Dept.\ of Physics \& Astronomy, Johns Hopkins University,  3400 N. Charles St., Baltimore, MD 21218, USA}


\email{Email: budavari@jhu.edu, basu.amitabh@jhu.edu}

\begin{abstract}

One of the outstanding challenges of cross-identification is multiplicity: detections in crowded regions of the sky are often linked to more than one candidate associations of similar likelihoods. We map the resulting maximum likelihood partitioning to the fundamental assignment problem of discrete mathematics and efficiently solve the two-way catalog-level matching in the realm of combinatorial optimization using the so-called Hungarian algorithm. We introduce the method, demonstrate its performance in a mock universe where the true associations are known, and discuss the applicability of the new procedure to large surveys.

\end{abstract}

\section{Motivation}
Modern astronomy surveys tenaciously observe the sky every night and their automated software pipelines produce a huge number of exposures. The datasets from separate telescopes and instruments are then routinely combined to enable all multicolor and time-domain studies. Over the last decade there has been a tremendous progress in the field of catalog matching to provide reliable resources for astronomical and cosmological measurements. \citet{BS08-BayesCrossID} introduced a framework based on Bayesian hypothesis testing that properly incorporated astrometric uncertainties, which yielded superior results \citep{heinis}, and proved to be flexible to accommodate a variety of scenarios from associating stars with unknown proper motions \citep{kerekes} or even models for radio morphology \citep{fan}. These methods are also paired with efficient indexing ideas and search algorithms to make the process of matching fast on the largest catalogs, e.g., \citep{htm, healpix, igloo, zones, gpumatch}

The methods before, however, were only concerned with assigning a reliable quality measure of the association for a given set of sources. The likelihood of the association would be compared to the likelihood of the detections belonging to separate objects using the Bayes Factor and the previously used approaches do not resolve situations where a single detection appears in multiple associations with similar likelihoods. Previous recommendations to deal with such situations resorted to including more data, e.g., photometric measurements to refine the likelihood measurements \citep{marquez}.
In other words all candidate associations so far have been considered in isolation ignoring the fact that a star, for example, could only appear in a single match. Such exclusion rules can potentially affect a large fraction of the catalog especially in crowded fields. 

In this paper we study catalog-level cross-identification to provide strategies for finding the most likely matched catalogs in which every association is valid and no detections appear in multiple matches.
In \csect{sec:catxid} we formulate the catalog-level matching problem and map the likelihood optimization onto discrete minimization that is solved efficiently by the so-called Hungarian algorithm.
\csect{sec:app} details our implementation and its application to realistic mock catalogs.
In \csect{sec:disc} we discuss the results and \csect{sec:sum} concludes the paper.

\begin{figure*}
\epsscale{1.2}
\plotone{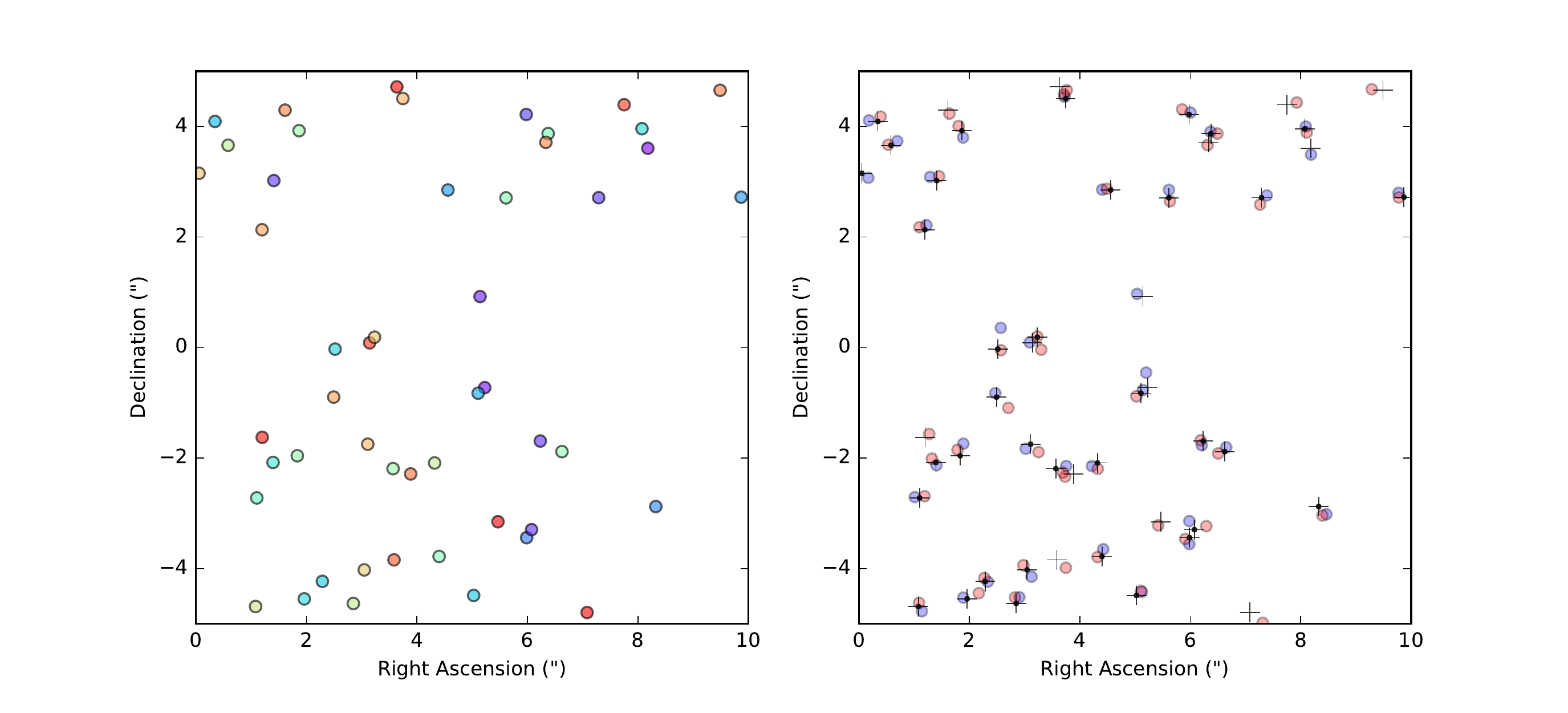}
\caption{\emph{Left:} A mock universe is constructed by randomly assigning coordinates (circles) and internal properties (color) to objects. \emph{Right:} Simulated surveys observe different subsets of the objects with overlap. The $+$ signs show the true directions of the objects but the \emph{red} and \emph{blue} dots illustrate the noisy measurements of the simulated sources. Objects that meet both selection functions are noted with small black dots on the true positions.
}
\label{fig:mock}
\end{figure*}

\section{Catalog-level Matching}
\label{sec:catxid}

Matching entire catalogs is a much more challenging task than calculating the evidence for a match. Based on the same equations one has to find the optimal configuration such that all matches in the output catalog are globally optimal. Before we turn to the mathematical formulation, we introduce the notion of partitioning \citep{budavari_loredo} and illustrate the problem in a simple scenario.

Let us consider two catalogs with 2 sources in catalog 1 and only 1 in catalog 2. In this situation there are only 3 possibilities: the source in catalog 2 is either matched to one of the two sources in catalog 1 or it is a completely separate one by itself.
If the pair $(i,c)$ denotes item $i$ in catalog $c$, we can write the possible partitions, excluding combinations of associations within each catalog, as
\begin{eqnarray}
\partn_0 &=& \Big\{\; \big\{ \textcolor{blue}{(1,1)} \big\}, 
                   \; \big\{ \textcolor{blue}{(2,1)} \big\}, 
                   \; \big\{ \textcolor{red}{(1,2)}  \big\} 
   	       \;\Big\} \nonumber \\   
\partn_1  &=& \Big\{\; \big\{ \textcolor{blue}{(1,1)} \big\}, 
	                \; \big\{ \textcolor{blue}{(2,1)}, 
                              \textcolor{red}{(1,2)} \big\} 
            \;\Big\}\nonumber \\                  
\partn_2 &=& \Big\{\; \big\{ \textcolor{blue}{(1,1)}, 
                             \textcolor{red}{(1,2)} \big\}, 
 	     	   	   \; \big\{ \textcolor{blue}{(2,1)} \big\} 
           \;\Big\} \nonumber
\end{eqnarray}
where $\partn_0$ corresponds to three separate objects but $\partn_1$ and $\partn_2$ represent alternative matching scenarios with only two objects. 
It is true that, in general, different partitioning schemes can have a different number of associations or objects. We introduce some notation to make the problem mathematically precise. Let $D$ denote the dataset of all $(i,c)$ pairs for all catalogs $c$ and items $i$ in catalog $c$. A {\em valid partition} $\partn$ is a set partition of $D$ into disjoint subsets $S_o \subseteq S$ indexed by $o\in O_\partn$ satisfying the condition that each set $S_o$ contains at most one item from any catalog. More precisely, for two distinct tuples $(i_1, c_1)$ and $(i_2, c_2)$ in $S_o$, we must have $c_1 \neq c_2$. For example, with two catalogs, each $S_o$ contains at most two tuples. An $S_o$ of size 1 will be called an {\em orphan} and $S_o$ with size 2 will be called an {\em association}. The collection of indices $o \in O_\partn$ will be called the set of {\em objects} of the valid partition $\partn$.


%
%

Our goal is to find the best valid partition that maximizes the likelihood function of the valid partitions. The fundamental equations of the Bayesian cross-identification are still applicable to the individual associations but here we consider all combinatorial possibilities.

\subsection{Likelihoods of Matched Catalogs}
\label{sec:likelihood}

The likelihood of a given valid partition $\partn$ is the probability density of the data \mbox{$D\!=\!\{(i,c)\}$} 
given valid partition $\partn$, and can be factored into the marginal likelihoods $\mlike_o$ of the objects $o \in O_\partn$ of the valid partition,
\be
\like(\partn) \equiv p(D|\partn) = \prod_{o\in O_\partn} \mlike_o
\label{like-partn}
\ee
The marginal likelihoods $\mlike_o$ are obtained by integrating the prior on the direction times the likelihood function over the entire sky \citep{budavari_loredo}: 
%
\be \mlike_o = \int d\drxn \; \;
\rho_{\cset(o)}(\drxn)\!\!\!\!\!\prod_{(i,c)\in \sset_o}\!\!\!\!\slike_{ic}(\drxn) \label{obj-like} \ee
where $\slike_{ic}(\drxn)$ is the likelihood function of source $i$ in catalog $c$ and $\rho_{\cset(o)}(\drxn)$ is the prior approriate for all catalogs $\cset(o)$ that contribute to object $o$.
For example, in the 2-way matching scenario there will only be up to two terms in the likelihood product that correspond to the two sources in the separate catalogs. The footprint of the catalogs along with the astrometric uncertainties determine the prior but in practice an isotropic prior is often an acceptable approximation for most associations when scaled \citep{BS08-BayesCrossID}.

The marginal likelihood can be analytically calculated for the \citet{fisher} distribution, \mbox{$\slike(\drxn)\!=\!f(\obs;\drxn,\kappa)$}, which is specified by two parameters: the concentration parameter $\kappa$, and a unit vector representing the direction of the mode, $\drxn$,
\be
f(\obs;\drxn,\kappa) = \frac{\kappa}{4\pi\,\sinh{}\kappa}\ \exp \Big( \kappa\,\drxn\obs \Big)
\ee
In the limit of small uncertainties typical in today's astronomy surveys, this distribution on the surface of the sphere can be well-approximated with the bivariate normal distribution on the tangent plan.

Next we apply the usual logarithmic transformation and, instead of maximizing the likelihood, we minimize 
\be\label{eq:obj}
A(\partn) \equiv -\ln \like(\partn) = -\sum_o \ln \mlike_o
\ee
When matching two catalogs, the terms in the sum correspond to pairs of sources in separate catalogs (defined as associations above), or orphan sources without a matching pair. 
We have a well-defined cost function to minimize by finding the best matching pairs. 
Similar problems have long been studied in discrete mathematics and algorithms are known to find the solution in polynomial time, that also run extremely efficiently in practice. We make a brief introduction of these ideas next, and show how to model the valid partition problem above using these ideas.

\begin{figure*}
\epsscale{1.2}
\plotone{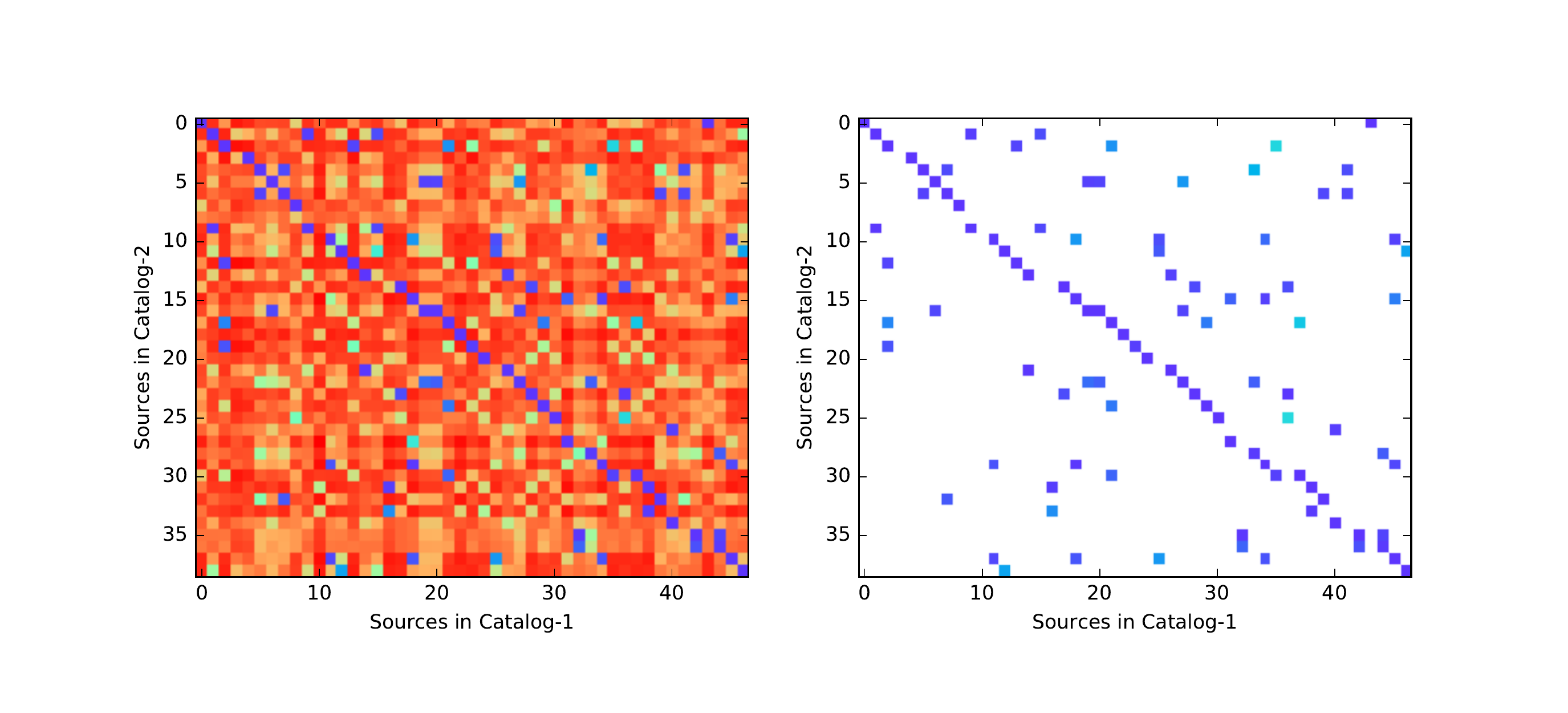}
\caption{The weight matrix $W$ that corresponds to the assignment problem contains the two catalogs along the axes. In the left panel, he \emph{bluer} matrix elements indicate more likely possible associations and the \emph{redder} ones are farther from each other.
In the right panel, we plot the sparsified weight matrix where the very unlikely candidates are omitted; see text.}
\label{fig:matrix}
\end{figure*}

\subsection{The Assignment Problem}

One can pose the problem of minimizing~\eqref{eq:obj} in the following equivalent way. Let the sizes of the two catalogs be $n_1$ and $n_2$. Introduce a \mbox{$n_1\!\times\!n_2$} matrix $W$, whose rows are indexed by the data points $(i,c)$ in the first catalog and the columns are indexed by the data points in the second catalog. 
%
%

The entry in the matrix $W$ corresponding to a row indexed by $(i_1, c_1)$ and a column indexed by $(i_2, c_2)$ (where of course, by definition, $c_1$ is the label for the first catalog and $c_2$ is the label for the second catalog), is given by the formula:

$$
-\ln \left(
    \int d\drxn \;\rho_{\{c_1,c_2\}}(\drxn)\;
    	\slike_{i_1c_1}(\drxn)\;\slike_{i_2c_2}(\drxn)
    \right)
$$
%

Now here is the key observation. Let $\mathcal{U}$ be the set of matrices of size \mbox{$n_1\!\times\!n_2$} with entries in $\{0,1\}$ such that every row and every column has at most one non zero entry. Any $P \in \mathcal{U}$ can be interpreted as a valid partition of $D$: For any $1$ entry in the matrix, the objects corresponding to the row and column with this entry are matched and form an association. The remaining objects are all orphans; equivalently, if any object has only $0$'s in its corresponding row or column then it is assigned as an orphan. 
Conversely, for any valid partition $\partn$ one can associate a matrix $P_\partn \in \mathcal{U}$ that has the property that every non-zero entry corresponds to an association. 
Therefore, there is a one-to-one correspondence between valid partitions and $\mathcal{U}$. One can then show the following:

\begin{theorem}
The valid partition $\partn$ that minimizes~\eqref{eq:obj} corresponds to a matrix $P\in \mathcal{U}$ that minimizes  
the trace $\Tr\,(W^T\!P)$ of the matrix $W^T\!P$. Thus, minimizing~\eqref{eq:obj} is equivalent to solving
\begin{equation}\label{eq:assignment}
\min\!\Big\{\!\Tr(W^T\!P): P\!\in\!\mathcal{U}\Big\} 
\end{equation}
\end{theorem}

Given any matrix \mbox{$W\!\in\!\mathbb{R}^{n\times{}n}$}, the {\em Hungarian Algorithm} \citep{Kuhn55thehungarian,Munkres57hunalg} is precisely a method to solve~\eqref{eq:assignment}. This is a classically studied combinatorial optimization problem called the {\em Assignment problem}. The Hungarian Algorithm that solves it, has a guaranteed running time of $\O{(n_1\!+\!n_2)^3}$
, but performs extremely fast in practice.

\begin{figure*}
\epsscale{1.2}
\plotone{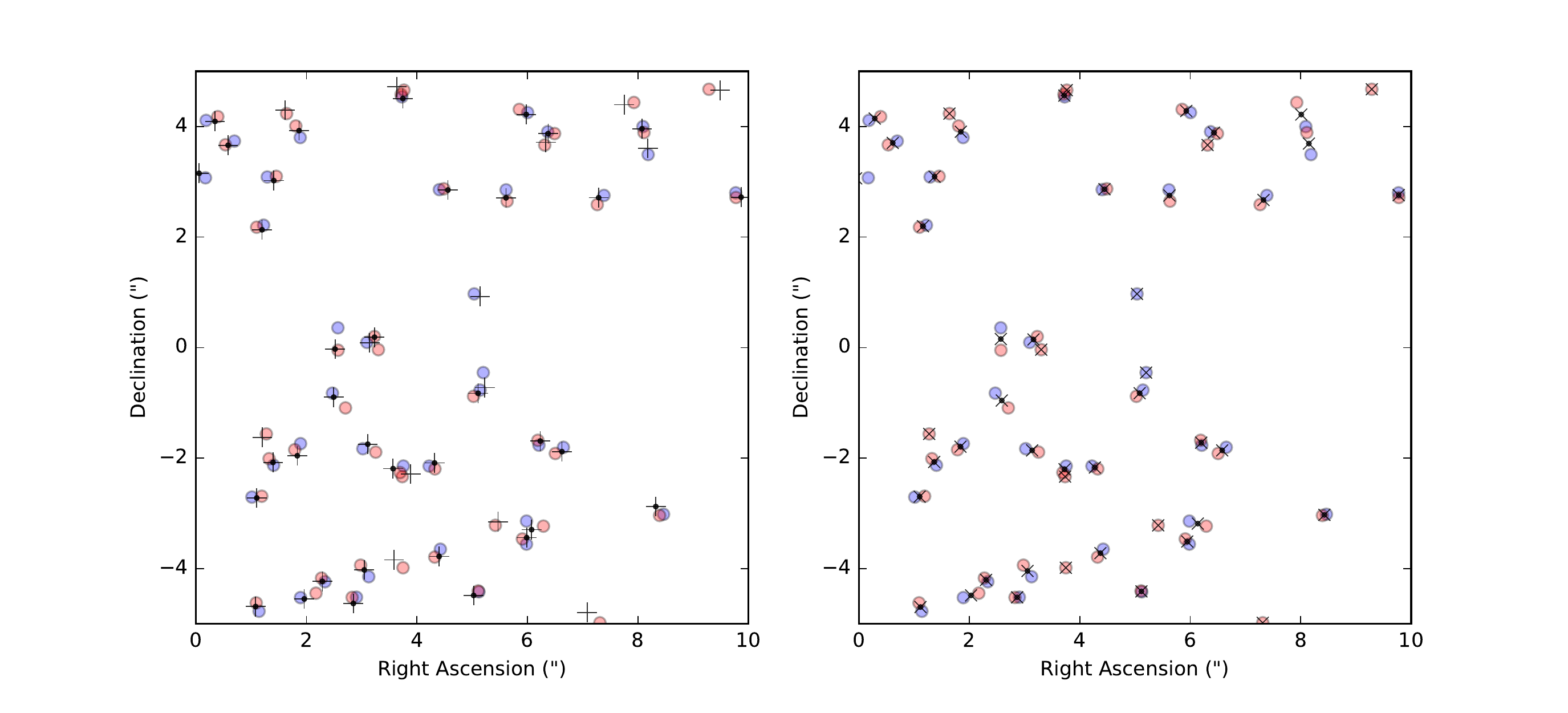}
\caption{The two panels compare the truth in the \emph{left} panel (same as in Figure~\ref{fig:mock}) and the optimal assignment of the detections on the \emph{right}, where the \emph{crosses} show the best-guess direction of the associations. The only wrong association is in the top right corner, see text.}
\label{fig:match}
\end{figure*}

\mbox{ }
\vspace{5mm}
\mbox{ }

\section{Application}
\label{sec:app}

The new probabilistic catalog-level cross-identification approach was tested on synthetic catalogs of a mock universe. In such settings we can include realistic artifacts that match typical observations but still know the ground truth.

\subsection{Mock Objects and Surveys}
Our mock universe contains point-like objects in a small field of view. 
In addition to a random direction $\obs$, each object has a  random property \mbox{$u_{01}\!\in\![0,1]$}, which could be thought of as the temperature of a star. 
The simulated sources are realizations of the objects with directions perturbed according to the chosen astrometric uncertainties. The selection effects, e.g., limiting magnitude, is determined by preset interval constraints on $u_{01}$. The surface density of a catalog can be adjusted by the length of the interval. The commonality of two catalogs is set by the overlap of the selection interval. \cfig{fig:mock} illustrates a mock universe with the colors showing the properties of the objects on the left, and two catalog realizations are shown on the right. 

The test scenario simulates catalogs in the flat-sky approximation using Gaussian with 
0.1-arcsecond astrometric uncertainties, which is similar to that of the Sloan Digital Sky Survey (SDSS). 
Using an isotropic prior the weight matrix $W$ is analytically calculated the same way as current procedures evaluate the Bayes factors.
\cfig{fig:matrix} illustrates the structure and the values of the weights. The left and right panels differ only by the hopelessly bad matches (shown in red on the left), which can be eliminating upfront before even considering the combinatorial problem. In practice we always work with such sparse matrices of the candidates (right panel), which can be quickly calculated by existing procedures. 
We note the ridge of good matches in the original \mbox{$n_1\!\times\!n_2$} matrix, which is an artifact of the simulations: sources appear in the same order as the mock objects. The advantage of this view is that we see the large number of confused elements that do not follow the linear trend. 


\cfig{fig:match} shows the ground truth (left) and the matched results (right). The $+$ signs illustrate the directions of the true objects as before and the result of our matching procedure using the $\times$ glyphs (right). The simulated sources are displayed as red and blue dots in both panels for easy comparison.
We see that even in this extremely confused field where the astrometric uncertainly would allow for multiple matches for almost all sources, the objects are perfectly resolved in almost all cases even in the most crowded parts of the field of view. The assignments correspond to objects with multiple detections and even the orphan sources are correctly isolated.
The only exceptions are due to selection effects, as illustrated in the top right corner of the field. In the left panel we see that in the mock universe there are three separate objects: the middle one is detected in both catalogs but the other two are only in one or the other. In this situation with the coordinates so close to each other, the algorithm finds that instead of matching the middle sources to each other, it is better to assign them to the nearby sources. While this is incorrect, we see that given the data, the output shown in the right panel is the optimal solution: having 2 pairs is better than 1 pair and 2 orphans.

\begin{figure}[b]
\epsscale{1.05}
\plotone{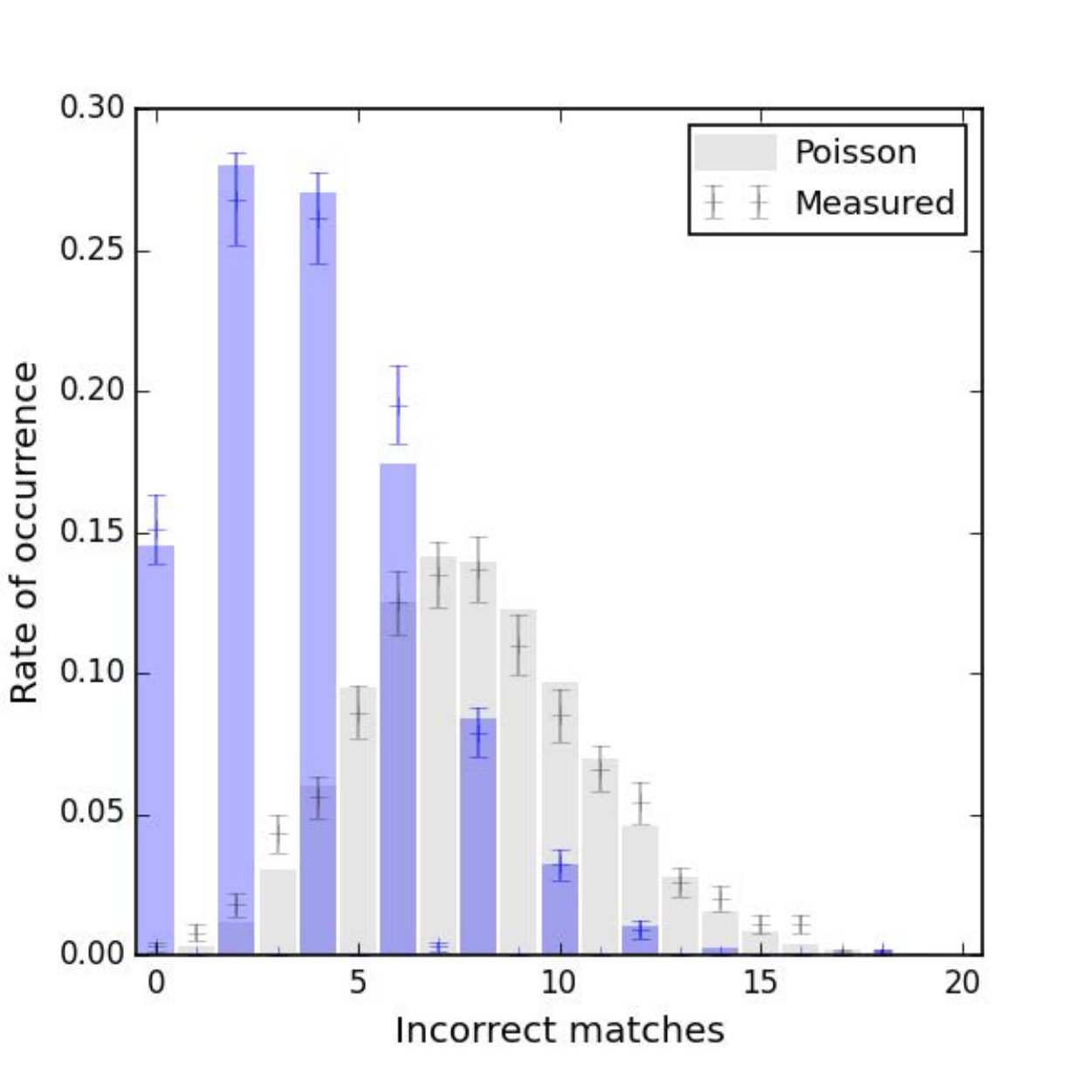}
\caption{The rate of incorrect outputs when matching source lists repeatedly. The gray points shows the number of incorrect matches for the nearest neighbor associations in a \mbox{3'$\times$3'} field of view with surface density of 400 objects per sq.~arcmin. The error bars simply represent the square root of the counts and the solid bars illustrate the Poisson distribution with $\lambda$ set to the sample mean. The blue points and error bars correspond to results from our new approach, see text.}
\label{fig:sim}
\end{figure}

\subsection{Evaluation of Repeated Simulations}
\label{sec:test}
Using the same simulation routines a suite of tests were performed to quantify the performance of the new approach in comparison to the traditional nearest neighbor methods, which in this simulation is equivalent to the maximum likelihood or Bayes factor matches due to the assumed homoscedastic astrometric errors.

For each mock we uniformly generate objects in a \mbox{3'$\times$3'} field of view with surface density of 400 per square arc minutes. 
Using an astrometric uncertainty of 0.04", 
two catalogs were generated including every object to best illustrate the effect of the exclusion rule. 

The simulated catalogs were matched by both finding the closest sources and by solving the assignment problem with the Hungarian algorithm.
\cfig{fig:sim} summarizes the distribution of the number of incorrect matches in the two scenarios. The gray error bars represent the measurements and error estimated from the square root of the counts.
We see that perfect matches practically never occur.
The Poisson distribution with the parameter set to the sample mean \mbox{($\lambda$=7.9)} is a close fit as shown by the solid bars. 

The blue error bars represent the new results. The first thing we notice is the gaps at odd numbers -- mistakes tend to occur in pairs due to the exclusion rules. The new method provides perfect matches 15\% of the time and the distribution is in general better skewed toward zero.
While the traditional method yields more than 4 incorrect matches with 90\% probability, the new method shrinks this to just 30\%.
The sample mean improves to 3.87 and the mode by a essentially a factor of 4.
The shape of the distribution can be actually well approximated by a Poisson distribution with parameter set to the half the sample mean \mbox{($\lambda$=1.9)} and stretched along the horizontal axis by a factor of two.

\section{Discussion}
\label{sec:disc}

The results that correspond to the maximum likelihood solution of the entire catalog appear exceptionally reliable and go beyond the limitations of working with individual associations. 
With the marginal likelihoods for the winning associations we can go back and derive the probability of each match using the prior information on surface density of the sources in the catalogs \citet{BS08-BayesCrossID}. 

\subsection{Computational Cost}

The proposed procedure is clearly more expensive than previous methods. 
As mentioned earlier, the general assignment problem solved by the Hungarian algorithm requires $\O{(n_1\!+\!n_2)^3}$ 
time, where $n_1$ and $n_2$ are the number of rows and columns of the input matrix $W$ (in our case \mbox{$|D|\!=\!n_1\!+\!n_2$}). 
For example, the 3600$\times$3600 weight matrices of our simulations were processed by the  solver called \texttt{linear\_sum\_assignment()} in the \texttt{optimize} module of {SciPy} \citep{scipy} in approximately 2.2 seconds on a desktop computer.
While this sounds prohibitively slow at first, when considering large astronomy surveys of hundreds of millions of sources, in practice the wall-clock time will be actually much shorter due to the sparsity of the weight matrix. In other words the list of candidate associations is actually only on the order of $\max(n_1,n_2)$ long instead of the number of pairs, which naively would be \O{n_1n_2}.

In fact resolving the multiplicity issue can be reserved for the overlapping associations where friends-of-friends connected components emerge as problem cases. Luckily these can be dealt with separately, hence reducing the cardinality of the problem to much smaller numbers of sources. \citet{hsc} used a greedy approach to find the best the partitions using clustering algorithms and Bayesian hypothesis testing. In the 2-way matching case, we can actually afford to do a global search for the optimum with existing methods such as the Hungarian algorithm.

\subsection{Likely Matched Catalogs}

Finding the optimal matched catalog is unfortunately not the final answer. There can be other realizations with the same or similarly good likelihoods. In principle these alternatives should be reported for further statistical analysis and marginalized over in the study. 

While the approach applied in our study does not automatically provide the sub-optimal solutions, the optimum would a good starting point for some Markov chain Monte Carlo algorithm to explore alternative matched catalogs. 
The cost of storing or automatically generating these catalogs and providing the  uncertainty estimates on the associations or their statistics needs to be studied further but is straightforward from the methodology's point of view.

\section{Summary}
\label{sec:sum}

Our method builds on previous developments in the field of cross-identification. The candidate associations are found by efficient matching procedures using ideas in database indexing and search trees and their sparse weight matrix is derived from the Bayesian approach introduced earlier. The conceptual change is to do a joint optimization to find the best matched catalog among the possible competing scenarios.

We have demonstrated that the catalog-level matching problem can be efficiently solved in case of two input datasets. 
The maximum of the marginal likelihood of the entire matched catalog is found using  discrete optimization such as the Hungarian algorithm in the case of our prototype. Despite the complexity of the problem, the sparse nature of rare \O{n} associations suggests that the new method is applicable to the largest catalogs.

Matching more than two catalogs, however, remains to be a challenge as no analog of the Hungarian algorithm or similar exists for the general case. The only hope is that the associations become even sparser and in that limit new algorithms can be developed.
Traditional matching procedures, which do not consider the multiplicity issue, can efficiently find candidate associations by adding catalogs one-by-one using a recursive formula 
\citep{BS08-BayesCrossID,budavari_skyquery:_2013}. Our future plans include examining the connectivity graph of the associations and work with the problem cases separately.

\section*{Acknowledgement}
The authors are grateful to Tom Loredo and Steve Lubow for invaluable discussions on various aspects of this study. 
TB acknowledges partial support from the NSF via grant AST-1412566 and NASA via the awards NNG16PJ23C and STSci-49721 under NAS5-26555.

\iftrue

\bibliography{XMATCH.bib}

\begin{thebibliography}{18}%
\makeatletter
\providecommand \@ifxundefined [1]{%
 \@ifx{#1\undefined}
}%
\providecommand \@ifnum [1]{%
 \ifnum #1\expandafter \@firstoftwo
 \else \expandafter \@secondoftwo
 \fi
}%
\providecommand \@ifx [1]{%
 \ifx #1\expandafter \@firstoftwo
 \else \expandafter \@secondoftwo
 \fi
}%
\providecommand \natexlab [1]{#1}%
\providecommand \enquote  [1]{``#1''}%
\providecommand \bibnamefont  [1]{#1}%
\providecommand \bibfnamefont [1]{#1}%
\providecommand \citenamefont [1]{#1}%
\providecommand \href@noop [0]{\@secondoftwo}%
\providecommand \href [0]{\begingroup \@sanitize@url \@href}%
\providecommand \@href[1]{\@@startlink{#1}\@@href}%
\providecommand \@@href[1]{\endgroup#1\@@endlink}%
\providecommand \@sanitize@url [0]{\catcode `\\12\catcode `\$12\catcode
  `\&12\catcode `\#12\catcode `\^12\catcode `\_12\catcode `\%12\relax}%
\providecommand \@@startlink[1]{}%
\providecommand \@@endlink[0]{}%
\providecommand \url  [0]{\begingroup\@sanitize@url \@url }%
\providecommand \@url [1]{\endgroup\@href {#1}{\urlprefix }}%
\providecommand \urlprefix  [0]{URL }%
\providecommand \Eprint [0]{\href }%
\providecommand \doibase [0]{http://dx.doi.org/}%
\providecommand \selectlanguage [0]{\@gobble}%
\providecommand \bibinfo  [0]{\@secondoftwo}%
\providecommand \bibfield  [0]{\@secondoftwo}%
\providecommand \translation [1]{[#1]}%
\providecommand \BibitemOpen [0]{}%
\providecommand \bibitemStop [0]{}%
\providecommand \bibitemNoStop [0]{.\EOS\space}%
\providecommand \EOS [0]{\spacefactor3000\relax}%
\providecommand \BibitemShut  [1]{\csname bibitem#1\endcsname}%
\let\auto@bib@innerbib\@empty
\bibitem [{\citenamefont {{Budav{\'a}ri}}\ and\ \citenamefont
  {{Szalay}}(2008)}]{BS08-BayesCrossID}%
  \BibitemOpen
  \bibfield  {author} {\bibinfo {author} {\bibfnamefont {T.}~\bibnamefont
  {{Budav{\'a}ri}}}\ and\ \bibinfo {author} {\bibfnamefont {A.~S.}\
  \bibnamefont {{Szalay}}},\ }\href {\doibase 10.1086/587156} {\bibfield
  {journal} {\bibinfo  {journal} {\apj}\ }\textbf {\bibinfo {volume} {679}},\
  \bibinfo {pages} {301} (\bibinfo {year} {2008})},\ \Eprint
  {http://arxiv.org/abs/0707.1611} {arXiv:0707.1611} \BibitemShut {NoStop}%
\bibitem [{\citenamefont {{Heinis}}\ \emph {et~al.}(2009)\citenamefont
  {{Heinis}}, \citenamefont {{Budav{\'a}ri}},\ and\ \citenamefont
  {{Szalay}}}]{heinis}%
  \BibitemOpen
  \bibfield  {author} {\bibinfo {author} {\bibfnamefont {S.}~\bibnamefont
  {{Heinis}}}, \bibinfo {author} {\bibfnamefont {T.}~\bibnamefont
  {{Budav{\'a}ri}}}, \ and\ \bibinfo {author} {\bibfnamefont {A.~S.}\
  \bibnamefont {{Szalay}}},\ }\href {\doibase 10.1088/0004-637X/705/1/739}
  {\bibfield  {journal} {\bibinfo  {journal} {\apj}\ }\textbf {\bibinfo
  {volume} {705}},\ \bibinfo {pages} {739} (\bibinfo {year} {2009})},\ \Eprint
  {http://arxiv.org/abs/0910.3214} {arXiv:0910.3214} \BibitemShut {NoStop}%
\bibitem [{\citenamefont {{Kerekes}}\ \emph {et~al.}(2010)\citenamefont
  {{Kerekes}}, \citenamefont {{Budav{\'a}ri}}, \citenamefont {{Csabai}},
  \citenamefont {{Connolly}},\ and\ \citenamefont {{Szalay}}}]{kerekes}%
  \BibitemOpen
  \bibfield  {author} {\bibinfo {author} {\bibfnamefont {G.}~\bibnamefont
  {{Kerekes}}}, \bibinfo {author} {\bibfnamefont {T.}~\bibnamefont
  {{Budav{\'a}ri}}}, \bibinfo {author} {\bibfnamefont {I.}~\bibnamefont
  {{Csabai}}}, \bibinfo {author} {\bibfnamefont {A.~J.}\ \bibnamefont
  {{Connolly}}}, \ and\ \bibinfo {author} {\bibfnamefont {A.~S.}\ \bibnamefont
  {{Szalay}}},\ }\href {\doibase 10.1088/0004-637X/719/1/59} {\bibfield
  {journal} {\bibinfo  {journal} {\apj}\ }\textbf {\bibinfo {volume} {719}},\
  \bibinfo {pages} {59} (\bibinfo {year} {2010})},\ \Eprint
  {http://arxiv.org/abs/1006.2096} {arXiv:1006.2096 [astro-ph.GA]} \BibitemShut
  {NoStop}%
\bibitem [{\citenamefont {{Fan}}\ \emph {et~al.}(2013)\citenamefont {{Fan}},
  \citenamefont {{Budav{\'a}ri}}, \citenamefont {{Szalay}}, \citenamefont
  {{Cui}},\ and\ \citenamefont {{Zhao}}}]{fan}%
  \BibitemOpen
  \bibfield  {author} {\bibinfo {author} {\bibfnamefont {D.}~\bibnamefont
  {{Fan}}}, \bibinfo {author} {\bibfnamefont {T.}~\bibnamefont
  {{Budav{\'a}ri}}}, \bibinfo {author} {\bibfnamefont {A.~S.}\ \bibnamefont
  {{Szalay}}}, \bibinfo {author} {\bibfnamefont {C.}~\bibnamefont {{Cui}}}, \
  and\ \bibinfo {author} {\bibfnamefont {Y.}~\bibnamefont {{Zhao}}},\ }\href
  {\doibase 10.1086/669707} {\bibfield  {journal} {\bibinfo  {journal} {\pasp}\
  }\textbf {\bibinfo {volume} {125}},\ \bibinfo {pages} {218} (\bibinfo {year}
  {2013})},\ \Eprint {http://arxiv.org/abs/1403.4358} {arXiv:1403.4358
  [astro-ph.IM]} \BibitemShut {NoStop}%
\bibitem [{\citenamefont {{Kunszt}}\ \emph {et~al.}(2001)\citenamefont
  {{Kunszt}}, \citenamefont {{Szalay}},\ and\ \citenamefont {{Thakar}}}]{htm}%
  \BibitemOpen
  \bibfield  {author} {\bibinfo {author} {\bibfnamefont {P.~Z.}\ \bibnamefont
  {{Kunszt}}}, \bibinfo {author} {\bibfnamefont {A.~S.}\ \bibnamefont
  {{Szalay}}}, \ and\ \bibinfo {author} {\bibfnamefont {A.~R.}\ \bibnamefont
  {{Thakar}}},\ }in\ \href {\doibase 10.1007/10849171_83} {\emph {\bibinfo
  {booktitle} {Mining the Sky}}},\ \bibinfo {editor} {edited by\ \bibinfo
  {editor} {\bibfnamefont {A.~J.}\ \bibnamefont {{Banday}}}, \bibinfo {editor}
  {\bibfnamefont {S.}~\bibnamefont {{Zaroubi}}}, \ and\ \bibinfo {editor}
  {\bibfnamefont {M.}~\bibnamefont {{Bartelmann}}}}\ (\bibinfo {year} {2001})\
  p.\ \bibinfo {pages} {631}\BibitemShut {NoStop}%
\bibitem [{\citenamefont {{G{\'o}rski}}\ \emph {et~al.}(2005)\citenamefont
  {{G{\'o}rski}}, \citenamefont {{Hivon}}, \citenamefont {{Banday}},
  \citenamefont {{Wandelt}}, \citenamefont {{Hansen}}, \citenamefont
  {{Reinecke}},\ and\ \citenamefont {{Bartelmann}}}]{healpix}%
  \BibitemOpen
  \bibfield  {author} {\bibinfo {author} {\bibfnamefont {K.~M.}\ \bibnamefont
  {{G{\'o}rski}}}, \bibinfo {author} {\bibfnamefont {E.}~\bibnamefont
  {{Hivon}}}, \bibinfo {author} {\bibfnamefont {A.~J.}\ \bibnamefont
  {{Banday}}}, \bibinfo {author} {\bibfnamefont {B.~D.}\ \bibnamefont
  {{Wandelt}}}, \bibinfo {author} {\bibfnamefont {F.~K.}\ \bibnamefont
  {{Hansen}}}, \bibinfo {author} {\bibfnamefont {M.}~\bibnamefont
  {{Reinecke}}}, \ and\ \bibinfo {author} {\bibfnamefont {M.}~\bibnamefont
  {{Bartelmann}}},\ }\href {\doibase 10.1086/427976} {\bibfield  {journal}
  {\bibinfo  {journal} {\apj}\ }\textbf {\bibinfo {volume} {622}},\ \bibinfo
  {pages} {759} (\bibinfo {year} {2005})},\ \Eprint
  {http://arxiv.org/abs/astro-ph/0409513} {astro-ph/0409513} \BibitemShut
  {NoStop}%
\bibitem [{\citenamefont {{Crittenden}}(2000)}]{igloo}%
  \BibitemOpen
  \bibfield  {author} {\bibinfo {author} {\bibfnamefont {R.~G.}\ \bibnamefont
  {{Crittenden}}},\ }\href@noop {} {\bibfield  {journal} {\bibinfo  {journal}
  {Astrophysical Letters and Communications}\ }\textbf {\bibinfo {volume}
  {37}},\ \bibinfo {pages} {377} (\bibinfo {year} {2000})},\ \Eprint
  {http://arxiv.org/abs/astro-ph/9811273} {astro-ph/9811273} \BibitemShut
  {NoStop}%
\bibitem [{\citenamefont {{Gray}}\ \emph {et~al.}(2007)\citenamefont {{Gray}},
  \citenamefont {{Nieto-Santisteban}},\ and\ \citenamefont {{Szalay}}}]{zones}%
  \BibitemOpen
  \bibfield  {author} {\bibinfo {author} {\bibfnamefont {J.}~\bibnamefont
  {{Gray}}}, \bibinfo {author} {\bibfnamefont {M.~A.}\ \bibnamefont
  {{Nieto-Santisteban}}}, \ and\ \bibinfo {author} {\bibfnamefont {A.~S.}\
  \bibnamefont {{Szalay}}},\ }\href@noop {} {\bibfield  {journal} {\bibinfo
  {journal} {eprint arXiv:cs/0701171}\ } (\bibinfo {year} {2007})},\ \Eprint
  {http://arxiv.org/abs/cs/0701171} {cs/0701171} \BibitemShut {NoStop}%
\bibitem [{\citenamefont {{Lee}}\ and\ \citenamefont
  {{Budav{\'a}ri}}(2013)}]{gpumatch}%
  \BibitemOpen
  \bibfield  {author} {\bibinfo {author} {\bibfnamefont {M.~A.}\ \bibnamefont
  {{Lee}}}\ and\ \bibinfo {author} {\bibfnamefont {T.}~\bibnamefont
  {{Budav{\'a}ri}}},\ }in\ \href@noop {} {\emph {\bibinfo {booktitle}
  {Astronomical Data Analysis Software and Systems XXII}}},\ \bibinfo {series}
  {Astronomical Society of the Pacific Conference Series}, Vol.\ \bibinfo
  {volume} {475},\ \bibinfo {editor} {edited by\ \bibinfo {editor}
  {\bibfnamefont {D.~N.}\ \bibnamefont {{Friedel}}}}\ (\bibinfo {year} {2013})\
  p.\ \bibinfo {pages} {235}\BibitemShut {NoStop}%
\bibitem [{\citenamefont {{Marquez}}\ \emph {et~al.}(2014)\citenamefont
  {{Marquez}}, \citenamefont {{Budav{\'a}ri}},\ and\ \citenamefont
  {{Sarro}}}]{marquez}%
  \BibitemOpen
  \bibfield  {author} {\bibinfo {author} {\bibfnamefont {M.~J.}\ \bibnamefont
  {{Marquez}}}, \bibinfo {author} {\bibfnamefont {T.}~\bibnamefont
  {{Budav{\'a}ri}}}, \ and\ \bibinfo {author} {\bibfnamefont {L.~M.}\
  \bibnamefont {{Sarro}}},\ }\href {\doibase 10.1051/0004-6361/201322625}
  {\bibfield  {journal} {\bibinfo  {journal} {\aap}\ }\textbf {\bibinfo
  {volume} {563}},\ \bibinfo {eid} {A14} (\bibinfo {year} {2014})}\BibitemShut
  {NoStop}%
\bibitem [{\citenamefont {Budav\'ari}\ and\ \citenamefont
  {Loredo}(2015)}]{budavari_loredo}%
  \BibitemOpen
  \bibfield  {author} {\bibinfo {author} {\bibfnamefont {T.}~\bibnamefont
  {Budav\'ari}}\ and\ \bibinfo {author} {\bibfnamefont {T.~J.}\ \bibnamefont
  {Loredo}},\ }\href {\doibase 10.1146/annurev-statistics-010814-020231}
  {\bibfield  {journal} {\bibinfo  {journal} {Annual Review of Statistics and
  Its Application}\ }\textbf {\bibinfo {volume} {2}},\ \bibinfo {pages} {113}
  (\bibinfo {year} {2015})},\ \Eprint
  {http://arxiv.org/abs/http://dx.doi.org/10.1146/annurev-statistics-010814-020231}
  {http://dx.doi.org/10.1146/annurev-statistics-010814-020231} \BibitemShut
  {NoStop}%
\bibitem [{\citenamefont {{Fisher}}(1953)}]{fisher}%
  \BibitemOpen
  \bibfield  {author} {\bibinfo {author} {\bibfnamefont {R.}~\bibnamefont
  {{Fisher}}},\ }\href {\doibase 10.1098/rspa.1953.0064} {\bibfield  {journal}
  {\bibinfo  {journal} {Proceedings of the Royal Society of London Series A}\
  }\textbf {\bibinfo {volume} {217}},\ \bibinfo {pages} {295} (\bibinfo {year}
  {1953})}\BibitemShut {NoStop}%
\bibitem [{\citenamefont {Kuhn}\ and\ \citenamefont
  {Yaw}(1955)}]{Kuhn55thehungarian}%
  \BibitemOpen
  \bibfield  {author} {\bibinfo {author} {\bibfnamefont {H.~W.}\ \bibnamefont
  {Kuhn}}\ and\ \bibinfo {author} {\bibfnamefont {B.}~\bibnamefont {Yaw}},\
  }\href@noop {} {\bibfield  {journal} {\bibinfo  {journal} {Naval Res. Logist.
  Quart}\ ,\ \bibinfo {pages} {83}} (\bibinfo {year} {1955})}\BibitemShut
  {NoStop}%
\bibitem [{\citenamefont {Munkres}(1957)}]{Munkres57hunalg}%
  \BibitemOpen
  \bibfield  {author} {\bibinfo {author} {\bibfnamefont {J.}~\bibnamefont
  {Munkres}},\ }\href@noop {} {\enquote {\bibinfo {title} {Algorithms for the
  assignment and transportation problems},}\ } (\bibinfo {year}
  {1957})\BibitemShut {NoStop}%
\bibitem [{\citenamefont {{Ivezic}}\ \emph {et~al.}(2008)\citenamefont
  {{Ivezic}}, \citenamefont {{Axelrod}}, \citenamefont {{Brandt}},
  \citenamefont {{Burke}}, \citenamefont {{Claver}}, \citenamefont
  {{Connolly}}, \citenamefont {{Cook}}, \citenamefont {{Gee}}, \citenamefont
  {{Gilmore}}, \citenamefont {{Jacoby}}, \citenamefont {{Jones}}, \citenamefont
  {{Kahn}}, \citenamefont {{Kantor}}, \citenamefont {{Krabbendam}},
  \citenamefont {{Lupton}}, \citenamefont {{Monet}}, \citenamefont {{Pinto}},
  \citenamefont {{Saha}}, \citenamefont {{Schalk}}, \citenamefont
  {{Schneider}}, \citenamefont {{Strauss}}, \citenamefont {{Stubbs}},
  \citenamefont {{Sweeney}}, \citenamefont {{Szalay}}, \citenamefont
  {{Thaler}}, \citenamefont {{Tyson}},\ and\ \citenamefont {{LSST
  Collaboration}}}]{2008SerAJ.176....1I}%
  \BibitemOpen
  \bibfield  {author} {\bibinfo {author} {\bibfnamefont {Z.}~\bibnamefont
  {{Ivezic}}}, \bibinfo {author} {\bibfnamefont {T.}~\bibnamefont {{Axelrod}}},
  \bibinfo {author} {\bibfnamefont {W.~N.}\ \bibnamefont {{Brandt}}}, \bibinfo
  {author} {\bibfnamefont {D.~L.}\ \bibnamefont {{Burke}}}, \bibinfo {author}
  {\bibfnamefont {C.~F.}\ \bibnamefont {{Claver}}}, \bibinfo {author}
  {\bibfnamefont {A.}~\bibnamefont {{Connolly}}}, \bibinfo {author}
  {\bibfnamefont {K.~H.}\ \bibnamefont {{Cook}}}, \bibinfo {author}
  {\bibfnamefont {P.}~\bibnamefont {{Gee}}}, \bibinfo {author} {\bibfnamefont
  {D.~K.}\ \bibnamefont {{Gilmore}}}, \bibinfo {author} {\bibfnamefont {S.~H.}\
  \bibnamefont {{Jacoby}}}, \bibinfo {author} {\bibfnamefont {R.~L.}\
  \bibnamefont {{Jones}}}, \bibinfo {author} {\bibfnamefont {S.~M.}\
  \bibnamefont {{Kahn}}}, \bibinfo {author} {\bibfnamefont {J.~P.}\
  \bibnamefont {{Kantor}}}, \bibinfo {author} {\bibfnamefont {V.~V.}\
  \bibnamefont {{Krabbendam}}}, \bibinfo {author} {\bibfnamefont {R.~H.}\
  \bibnamefont {{Lupton}}}, \bibinfo {author} {\bibfnamefont {D.~G.}\
  \bibnamefont {{Monet}}}, \bibinfo {author} {\bibfnamefont {P.~A.}\
  \bibnamefont {{Pinto}}}, \bibinfo {author} {\bibfnamefont {A.}~\bibnamefont
  {{Saha}}}, \bibinfo {author} {\bibfnamefont {T.~L.}\ \bibnamefont
  {{Schalk}}}, \bibinfo {author} {\bibfnamefont {D.~P.}\ \bibnamefont
  {{Schneider}}}, \bibinfo {author} {\bibfnamefont {M.~A.}\ \bibnamefont
  {{Strauss}}}, \bibinfo {author} {\bibfnamefont {C.~W.}\ \bibnamefont
  {{Stubbs}}}, \bibinfo {author} {\bibfnamefont {D.}~\bibnamefont {{Sweeney}}},
  \bibinfo {author} {\bibfnamefont {A.}~\bibnamefont {{Szalay}}}, \bibinfo
  {author} {\bibfnamefont {J.~J.}\ \bibnamefont {{Thaler}}}, \bibinfo {author}
  {\bibfnamefont {J.~A.}\ \bibnamefont {{Tyson}}}, \ and\ \bibinfo {author}
  {\bibnamefont {{LSST Collaboration}}},\ }\href {\doibase 10.2298/SAJ0876001I}
  {\bibfield  {journal} {\bibinfo  {journal} {Serbian Astronomical Journal}\
  }\textbf {\bibinfo {volume} {176}},\ \bibinfo {pages} {1} (\bibinfo {year}
  {2008})}\BibitemShut {NoStop}%
\bibitem [{\citenamefont {Jones}\ \emph {et~al.}(2001)\citenamefont {Jones},
  \citenamefont {Oliphant}, \citenamefont {Peterson} \emph {et~al.}}]{scipy}%
  \BibitemOpen
  \bibfield  {author} {\bibinfo {author} {\bibfnamefont {E.}~\bibnamefont
  {Jones}}, \bibinfo {author} {\bibfnamefont {T.}~\bibnamefont {Oliphant}},
  \bibinfo {author} {\bibfnamefont {P.}~\bibnamefont {Peterson}},  \emph
  {et~al.},\ }\href {http://www.scipy.org/} {\enquote {\bibinfo {title}
  {{SciPy}: Open source scientific tools for {Python}},}\ } (\bibinfo {year}
  {2001}),\ \bibinfo {note} {[Online; accessed 2016-05-16]}\BibitemShut
  {NoStop}%
\bibitem [{\citenamefont {{Budav{\'a}ri}}\ and\ \citenamefont
  {{Lubow}}(2012)}]{hsc}%
  \BibitemOpen
  \bibfield  {author} {\bibinfo {author} {\bibfnamefont {T.}~\bibnamefont
  {{Budav{\'a}ri}}}\ and\ \bibinfo {author} {\bibfnamefont {S.~H.}\
  \bibnamefont {{Lubow}}},\ }\href {\doibase 10.1088/0004-637X/761/2/188}
  {\bibfield  {journal} {\bibinfo  {journal} {\apj}\ }\textbf {\bibinfo
  {volume} {761}},\ \bibinfo {eid} {188} (\bibinfo {year} {2012})},\ \Eprint
  {http://arxiv.org/abs/1206.0644} {arXiv:1206.0644 [astro-ph.IM]} \BibitemShut
  {NoStop}%
\bibitem [{\citenamefont {Budav\'ari}\ \emph {et~al.}(2013)\citenamefont
  {Budav\'ari}, \citenamefont {Dobos},\ and\ \citenamefont
  {Szalay}}]{budavari_skyquery:_2013}%
  \BibitemOpen
  \bibfield  {author} {\bibinfo {author} {\bibfnamefont {T.}~\bibnamefont
  {Budav\'ari}}, \bibinfo {author} {\bibfnamefont {L.}~\bibnamefont {Dobos}}, \
  and\ \bibinfo {author} {\bibfnamefont {A.~S.}\ \bibnamefont {Szalay}},\
  }\href {\doibase 10.1109/MCSE.2013.41} {\bibfield  {journal} {\bibinfo
  {journal} {Computing in Science \& Engineering}\ }\textbf {\bibinfo {volume}
  {15}},\ \bibinfo {pages} {12} (\bibinfo {year} {2013})}\BibitemShut {NoStop}%
\end{thebibliography}%


\begin{thebibliography}{}

\bibitem[Budav{\'a}ri 
\& Szalay(2008)]{pxid} Budav{\'a}ri, T., \& Szalay, A.~S.\ 2008, \apj, 679, 301 
 
\bibitem[Budav{\'a}ri et al.(2003)]{pxidr} Budav{\'a}ri, T., 
Connolly, A.~J., Szalay, A.~S., et al.\ 2003, \apj, 595, 59 

\bibitem[Budav{\'a}ri \& Loredo(2015)]{budavariloredo} Budav{\'a}ri, T., \& Loredo, T. J.\ 2015, Annual Review of Statistics and Its Application, 2, 113
 
\bibitem[Peebles(1980)]{peebles} Peebles, P.~J.~E.\ 1980, 
Research supported by the National Science Foundation.~Princeton, N.J., 
Princeton University Press, 1980.~435 p.,  

\bibitem[Rohde et al.(2006)]{rohde2006} Rohde, D.~J., Gallagher, 
M.~R., Drinkwater, M.~J., \& Pimbblet, K.~A.\ 2006, \mnras, 369, 2 

\bibitem[Sutherland 
\& Saunders(1992)]{sutherlandandsaunders} Sutherland, W., \& Saunders, W.\ 1992, \mnras, 259, 413 
 
\bibitem[York et al.(2000)]{york00} York, D.~G., Adelman, J., 
Anderson, J.~E., Jr., et al.\ 2000, \aj, 120, 1579 

\bibitem[Hogg(2001)]{2001AJ....121.1207H} Hogg, D.~W.\ 2001, \aj, 121, 1207 


\bibitem[Adler(1981)]{adler} Adler, R.~J.\ 1981, The Geometry of Random Fields, Chichester: Wiley, 1981,  

\bibitem[Bardeen et al.(1986)]{bbks} Bardeen, J.~M., Bond, J.~R., Kaiser, N., \& Szalay, A.~S.\ 1986, \apj, 304, 15 

\bibitem[Bond \& Efstathiou(1987)]{bond} Bond, J.~R., \& Efstathiou, G.\ 1987, \mnras, 226, 655

\bibitem[Budav{\'a}ri \& Szalay(2008)]{pxid} Budav{\'a}ri, T., \& Szalay, A.~S.\ 2008, \apj, 679, 301

\bibitem[Budav{\'a}ri(2011)]{2011ApJ...736..155B} Budav{\'a}ri, T.\ 2011, 
\apj, 736, 155 

\bibitem[Gregory \& Loredo(1992)]{gregory} Gregory, P.~C., \& Loredo, T.~J.\ 1992, \apj, 398, 146


\bibitem[Kaiser(2004)]{kaiser} Kaiser, N.\ 2004, ``The Likelihood of Point Sources in Pixellated Images'', Pan-STARRS internal report, PSDC-002-010-xx

\bibitem[Kessler et al.(2009)]{snana} Kessler, R., Bernstein, J.~P., Cinabro, D., et al.\ 2009, \pasp, 121, 1028 

\bibitem[Loredo(2012)]{loredo} Loredo, T.~J.\ 2012, arXiv:1206.4278 

\bibitem[Lund \& Rudemo(2000)]{lund} Lund, J., \& Rudemo, M.\ 2000, Biometrika, 87, 2, pp.235-249 (http://www.jstor.org/stable/2673461)

\bibitem[Kerekes et al.(2010)]{pmxid} Kerekes, G., Budav{\'a}ri, T., Csabai, I., Connolly, A.~J., \& Szalay, A.~S.\ 2010, \apj, 719, 59 

\bibitem[Press(1997)]{press} Press, W.~H.\ 1997, Unsolved Problems in Astrophysics, p.49-60, arXiv:astro-ph/9604126

\bibitem[Riess et al.(1995)]{riess95} Riess, A.~G., Press, W.~H., \& Kirshner, R.~P.\ 1995, \apjl, 438, L17 

\bibitem[Szalay et al.(1999)]{chisq} Szalay, A.~S., Connolly, A.~J., \& Szokoly, G.~P.\ 1999, \aj, 117, 68 

\end{thebibliography}

\else

\fi

\end{document}